\documentclass[graybox]{svmult}

\usepackage{type1cm}        % activate if the above 3 fonts are
\usepackage{makeidx}         % allows index generation
\usepackage{graphicx}        % standard LaTeX graphics tool
\usepackage{multicol}        % used for the two-column index
\usepackage[bottom]{footmisc}% places footnotes at page bottom
\usepackage{braket}

\usepackage{newtxtext}       % 
\usepackage[varvw]{newtxmath}       % selects Times Roman as basic font

\makeindex             % used for the subject index
\begin{document}

\title*{Quantum solutions to possible challenges of Blockchain technology}
\author{Nivedita Dey, Mrityunjay Ghosh and Amlan Chakrabarti
}
\institute{Nivedita Dey \at QRDLab, University of Calcutta \email{mail.dnivedita@gmail.com}
\and Mrityunjay Ghosh \at HCL Technologies, University of Calcutta \email{g.mrityunjay@gmail.com}
\and Amlan Chakrabarti \at University of Calcutta \email{acakcs@caluniv.ac.in}}
\maketitle

\abstract*{}

\abstract{Technological advancements of Blockchain and other Distributed Ledger Techniques (DLTs) promise to provide significant advantages to applications seeking transparency, redundancy, and accountability. Actual adoption of these emerging technologies requires incorporating cost-effective, fast, QoS-enabled, secure, and scalable design. With the recent advent of quantum computing, the security of current blockchain cryptosystems can be compromised to a greater extent. Quantum algorithms like Shor’s large integer factorization algorithm and Grover’s unstructured database search algorithm can provide exponential and quadratic speedup, respectively, in contrast to their classical counterpart. This can put threats on both public-key cryptosystems and hash functions, which necessarily demands to migrate from classical cryptography to quantum-secure cryptography. Moreover, the computational latency of blockchain platforms causes slow transaction speed, so quantum computing principles might provide significant speedup and scalability in transaction processing and accelerating the mining process. For such purpose, this article first studies current and future classical state-of-the-art blockchain scalability and security primitives. The relevant quantum-safe blockchain cryptosystem initiatives which have been taken by Bitcoin, Ethereum, Corda, etc. are stated and compared with respect to key sizes, hash length, execution time, computational overhead, and energy efficiency. Post Quantum Cryptographic algorithms like Code-based, Lattice-based, Multivariate-based, and other schemes are not well suited for classical blockchain technology due to several disadvantages in practical implementation. Decryption latency, massive consumption of computational resources, and increased key size are few challenges that can hinder blockchain performance. We aim to provide the different ways where quantum-aided solutions can overcome the challenges in blockchain technology and enhance scalability, security, and performance in the post-quantum era. Though quantum information itself is in the Noisy Intermediate Scale Quantum (NISQ) era and not yet developed to a high technology readiness level, this paper also covers the current challenges and future research direction in the near-term, practical realization of post-quantum blockchain.}

\section{Introduction}

Quantum computing is no longer just a theoretical concept but a palpable for enterprises and have a great potential in delivering business values by solving complex problems efficiently. Conventional security in IT systems is provided by several asymmetric key and symmetric key cryptographic algorithms.\cite{79} Asymmetric key cryptographic algorithms like RSA (Rivest Shamir Adleman) encryption, ECDSA (Elliptic Curve Digital Signature Algorithm), ECDH (Elliptic Curve Diffe Hellman) algorithm, EdDSA (Edwards-curve Digital Signature Algorithm) use hard problems in mathematics like discrete logarithm problem, large integer factorization problem as an underlying cryptographic security provider.\cite{77}\cite{5}\cite{6} On the other hand, symmetric key or shared key cryptographic algorithms like DES (Data Encryption Standard), TDES (Triple DES), AES (Advanced Encryption Standard) use block cipher techniques to generate a symmetric key for both encryption and decryption.\cite{79}

Blockchain and other Distributed Ledger Techniques (DLTs) are novel computational data structures that evolve significantly in the last years due to their ability to provide secure communications, data privacy, resilience, and transparency. Blockchain technologies like Bitcoin, Ethereum, Monero, Ripplecoin, Zcash, etc., can exhibit redundancy, transparency, and accountability through the use of public-key cryptography and hash functions.\cite{2} Blockchain users leverage public key cryptographic primitives to authenticate transactions.\cite{1} On the other hand, hash functions are introduced to generate digital signatures and create blocks that do not necessarily trust each other.

The unprecedented growth of quantum computing can pose potential threats to classical blockchain cryptographic primitives. Shor’s factorization algorithm can break asymmetric encryption with twice as many logical qubits as the key size, which necessarily means to achieve a 112-bit security level using a 2048-bit RSA key, a 4096-qubit quantum computer is sufficient enough. Similarly, on the other hand, breaking AES-128 encryption key will be eventually reduced to breaking a 64-bit symmetric encryption key. This is achieved by Grover’s unstructured database search algorithm, which can reduce the brute force attack time to its square root. \cite{maindoc} \cite{10} Possible quantum threats in the near-term quantum era lead to the development of some quantum-resistant cryptographic algorithms which quantum computers can not break in the near future. Post-Quantum Cryptography (PQC) deals with devising quantum-resistant algorithms. Some of the most promising PQC areas include hash-based cryptography, code-based cryptography, lattice-based cryptography, and multivariate-quadratic-equations cryptography. \cite{maindoc} In this article, we analyze both the scalability and security issues associated with classical blockchain.

We discuss first different scalability aspects of blockchain both from an existing standpoint and future standpoint. Consequently, it is shown that how quantum advantage can be exploited to solve scalability issues in blockchain techniques. In order to analyze the same, we have identified some research use cases from blockchain-aided fintech applications like supercharged data analyses, speed of calculation, reduction in the number of false-positive cases in fraud detection, and Monte-Carlo simulation where quantum computation can be applied.

Additionally, to guide researchers on the development of quantum-resistant blockchain applications, we first provide a broad view on potential threats that can be offered by Shor’s and Grover’s algorithm on current cryptographic primitives used by blockchain. Furthermore, the most relevant post-quantum cryptographic initiatives for generic quantum-resistant cryptography and specific post-quantum initiatives on the blockchain are analyzed. Then, we discuss the vulnerabilities of five blockchain cryptocurrencies like Bitcoin, Ethereum, Litecoin, Monero, and Zcash. We also study the feasibility of quantum attacks on these DLTs.  Since there are challenges associated with post-quantum blockchain, we end up our article by discussing future research direction towards the development of reliable blockchain security primitives.

\section{Blockchain and Distributed Ledger Techniques }

\begin{figure}
  \includegraphics[width=12cm , height=7cm]{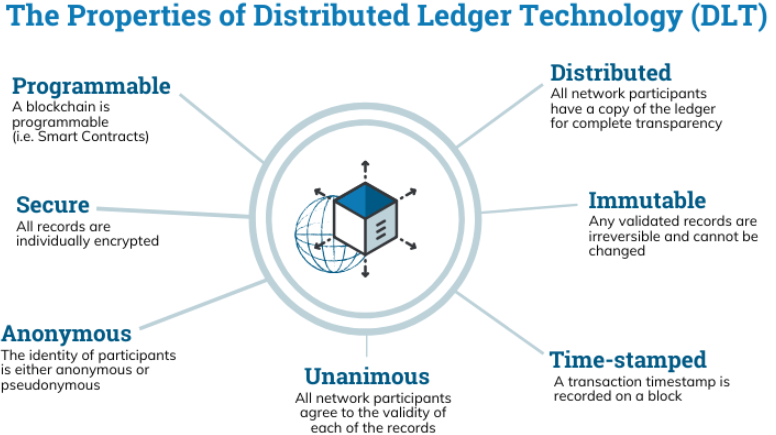}
  \caption{The properties of distributed ledger technology \cite{pic2}}
  \label{fig:dlt-properties}
\end{figure}

Blockchain or any DLT is a database that is consensually shared, replicated, and synchronized. The term ‘distributed ledger’ is associated with replicating and storing transaction data by each node or party on a Blockchain network. Coherence-related issues or database conflicts are handled automatically with pre-defined distributed ledger rules. Properties of a distributed ledger include operation between peer-to-peer networks, decentralization in keeping transaction data, allowance to consensus-based transactions, and tamper resistance. This makes a DLT to be programmable, secure, anonymous, unanimous, distributed, time-stamped, and immutable as shown in figure \ref{fig:dlt-properties}. The core parts of a blockchain consist of a block, a chain, a network, and a blockchain consensus mechanism.

\begin{figure}
  \includegraphics[width=12cm , height=7cm]{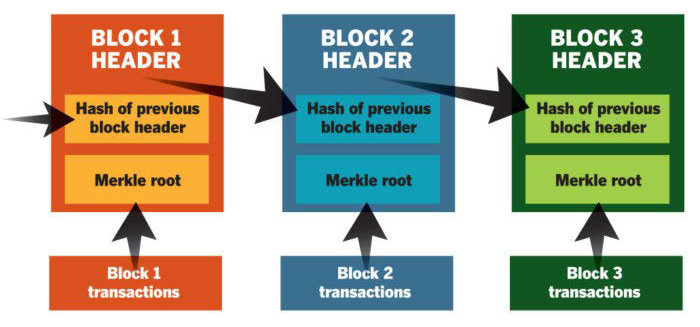}
  \caption{Blockchain technology in Bitcoin crypto-currency \cite{pic1}}
  \label{fig:bitcoinBlockchain}
\end{figure}

\begin{center}
\begin{table}
\begin{tabular}{p{3cm}|p{9cm}}
\hline
Variable Text \newline String Entry Value & Hash Value  \\
\hline
National Archives & 6429799b9af2d91cbf915cb0290f3a50281193a977b3457d63e4541cc5788c5b \\
\hline
National \ Archives \newline (an extra space) & d926fe7e72d09b249701dbcde2dad0ccb9b4bb653e053e461a67bbb951dcae0b \\
\hline
Nati0nal Archives & 5f2d570fc940d5f8de89310db43f789fdd99f51e89c021e1a50acb7a6fe2cf83 \\
\hline

\end{tabular}
\caption{Secure Hash Algorithm generating a 256-bit signature (SHA-256) generates different hash values for slight variations in the spelling of the words “National Archives”.}
\label{table:0}
\end{table}
\end{center}

\begin{center}
\begin{table}
\begin{tabular}{p{2cm}|p{10cm}}
\hline
Mechanism  & Operational Principle  \\
\hline
\hline
Proof of Work  & Miners solve complicated mathematical puzzles to receive a block as reward. Difficulty level for puzzles is determined by the mining speed.  \\
\hline
Proof of Stake  & Uses randomized process for selecting the node who will get chance to add a block. The user owning the biggest stake or owning coins for the longest time is more probable to create a new block. It is more energy-efficient approach than Proof of Work.  \\
\hline\\
Delegated Proof of Stake  & Uses stake-weighted voting system where users stake their coins to vote for a particular number of delegates. The delegate with highest number of votes is favoured to create a new block. It is one of the fastest consensus mechanisms for blockchain.  \\
\hline\\
Proof of \newline Capacity  & Users use digital storage to store the solutions for cryptographic puzzles. User with minimum storage capacity gets the chance to create a new block.  \\
\hline\\
Proof of Elapsed Time  & Uses amount of time spent by the users to choose producer of the new block. Each user is assigned with a random waiting time. The user with minimum waiting time gets chance to create the next block.  \\
\hline\\
Proof of \newline Authority  & Personal identities of the validators of the network are kept at stake. Only validator nodes are allowed to create new blocks. It is a modified version of Proof of Stake.  \\
\hline\\
Proof of Activity  & It is a combination of Proof of Work (blocks with mining reward address and header information) and Proof of Stake (use of header information to choose validators randomly).   \\
\hline\\
Proof of Identity  & Cryptographic evidence to ensure integrity and authenticity of created data by attaching an user's private key to a given transaction. Only an identified user can create a block in the network.  \\
\hline

\end{tabular}
\caption{Blockchain consensus mechanisms}
\label{table:first}
\end{table}
\end{center}

\begin{figure}
  \includegraphics[width=12cm , height=7cm]{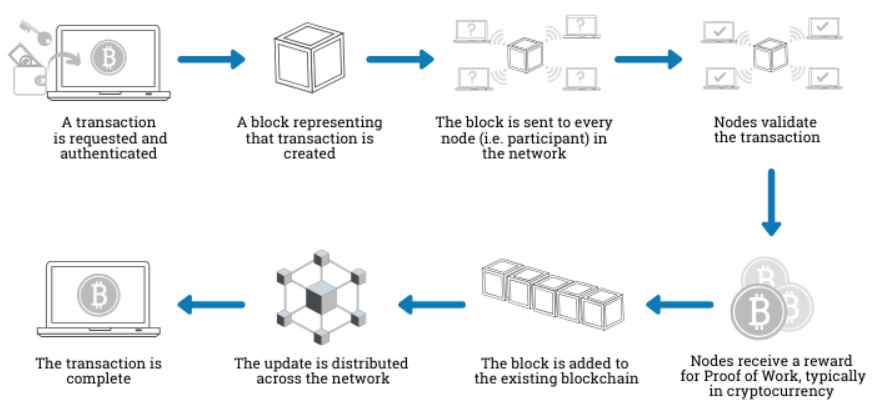}
  \caption{Flow of a transaction in Blockchain technology \cite{pic2}}
  \label{fig:transactionFlow}
\end{figure}
Block contains a list of transactions, which have been recorded periodically. Transactions represent any type of virtual activity taking place in a blockchain database. Block–specific rules are associated with the block itself at the time of network creation. Each block can set its size or the maximum number of transactions allowed to be stored within a block according to pre-defined rules.
\newline
The chaining of blocks is necessary whenever a block reaches its maximum limit in terms of the number of transactions. Then chaining or linking of blocks is done through hashing to connect the preceding block with the next block. The hash value of the preceding is inserted into the next block. Thus each block has an impact on the next block through cryptographic hashing. This has been shown in figure \ref{fig:bitcoinBlockchain}. Hash is an algorithm that takes a variable string of data as input and produces a fixed-length value as output. This hashing scheme is vital to generate small transaction data that can be easily validated and distributed to other nodes. If a block of data remains unaltered, then repeating a hash function on that block will always generate the same hash output. This can be further understood from table \ref{table:0}, where slight variations in the spelling of the word “National Archives” have led to different output values when a Secure Hash Algorithm (SHA – 256) is applied. Merkle tree is an extension of a simple hash algorithm where multiple hash values. The single hashed output, known as Merkle root, can represent multiple hashes. \cite{41}
\newline
A Blockchain network is made up of nodes or blocks, each containing a record of all the transactions. Each node in a Blockchain network is equiprobable to be trusted, and once a node solves a cryptographic puzzle (a mathematically hard problem ), it is allowed to add the next block to the Blockchain along with a performance reward. This technique is known as mining, and the node to solve the puzzle is known as miners.  \cite{44}
\newline
The blockchain consensus mechanism is the set of rules that govern the process of validating transactions to be entered into the Blockchain by the nodes, as shown in figure \ref{fig:transactionFlow}. These rules are established at the time of Blockchain creation. In case of a situation when both the parties want to initiate communication without knowing or trusting each other, a consensus mechanism is embedded to agree upon entry of transactions into the Blockchain. This can resolve Byzantine Generals Problem.\cite{12} Each Blockchain differs in its terms of consensus mechanism based on the type of active transactions it works with to facilitate authenticity and immutability of transaction records. Some well–known consensus mechanisms are Proof of Work (PoW), Proof of Stake (PoS), and delegated Proof of Stake (dPoS). \cite{maindoc} Other consensus mechanisms include Proof of Capacity, Proof of Elapsed Time, Proof of Authority, Proof of Activity, and Proof of Identity. Table \ref{table:first} represents the basic operational principles of all these consensus mechanisms. The process of finding a single reliable consensus mechanism is increasingly complex due to the nascency of the underlying platform. But, integrating Blockchain technology with business applications requires information about all these mechanisms.

\section{Scalability Aspects of Classical Blockchain}
 The thrive for achieving scalability in a blockchain is limitless, as adoption of blockchain is often bottlenecked due to the lack of a scalable solution. Bitcoin can process 4.6 TPS (Transactions Per Second); another competitor, Visa, processes 1,700 TPS on an average from over 150 million TPD (Transactions Per Day).\cite{1}
According to Deloitte Insights, “Blockchain-based systems are comparatively slow. Blockchain’s sluggish transaction speed is a major concern for enterprises that depend on high-performance legacy transaction processing systems”. Just to add a number for the percentage of adoption of blockchain by organizations, it is close to around 44\% based on a study published by Tata Communications. 
\subsection{Blockchain Scalability from Existing Standpoint}
Scalability issues in blockchain can account for uncontrolled transfer delays, high fees on the Bitcoin network, and congestion-uncontrolled traffic for the Ethereum blockchain network on which thousands of decentralized transactions are operated.\cite{60}\cite{76}

\subsubsection{Blockchain Scalability Metric} Blockchain transaction speed is a function of average block size and average transaction size. Mathematically, the transaction speed of a blockchain network can be visualized as, 
Number of Transactions per Block = Average Block Size in Bytes/ Average Transaction Size in Bytes 
\subsubsection{Market Incompetency of Bitcoin TPS} Block size of Bitcoin is hard-cored at 1 million bytes (1,048,576 bytes), and its average transaction size is 380.04 bytes. This leads to around 2759.12 transactions that can fit in one of Bitcoin’s blocks and approximately providing a TPS of only 4.6. Prior attempts were made by Bitcoin community to increase the scalability by changing parameters in the blockchain. For this purpose, the metrics chosen were block size (in bytes, to be increased to increase TPS) and block generation time (in seconds, to be adjusted by changing the complexity of the hashing puzzle to reduce its value). To achieve speedup in TPS like Visa, Bitcoin would require to increase its block size to 377.5 million bytes, which is 377.5x of default block size in Bitcoin.\cite{1} Keeping the block generation time of 10 minutes and average transaction size of 380 bytes(same as default) could ideally give a TPS of 1736. But, practically, this speedup can not be achieved as propagation delay or relay time will be required to broadcast a new block to every other node on the Bitcoin network. 
In a Bitcoin network comprising of 10,198 nodes, approximately 14 seconds are required to transmit 99\% of the block size of 1 million bytes across peers. Alternatively, if block generation time is less than 14 seconds, a new block will be generated before an old block gets received by most of the existing blocks. 
\subsubsection{SegWit – Flow Control in Blocks of Blockchain} 
SegWit or Segregated Witness aims at providing a solution to scalability issues of the Bitcoin network. It believes in segregating the witness part of each transaction from the actual transaction. The actual transaction is the base transaction data that covers how bitcoins are moving, where they are being moved to, and how they are being accessed. But, the “witness” part comprises a bit of code with cryptographic signature data to prove that the owner of bitcoin is interested in spending the bitcoin. This prevents the denial-of-service actions and helps to achieve non-repudiation. The soft fork of SegWit approach helps to improve block size without changes to core code, but the real challenge of scalability is in blockchain value propositions. The high propagation time of transactions can lead to significant security issues like double-spend attacks.
\newline
\textbf{Double-Spend Attack in Blockchain:}
The computational power of a decentralized proof-of-work system is the cumulative sum of the individual computational power of the nodes. More considerable computational power increases the probability of winning the mining reward for each newly created block, thus making an incentive or biasedness among nodes in mining pools. If a collection can achieve 51\% hashing power, it can effectively overturn network transactions and create a double-spend attack. Bitcoin fork ‘Bitcoin Gold’ (BTG) was hit by such an attack in 2018 and again in 2020 at the cost of around \$72,000 worth of BTG.\cite{78}
\subsection{Future Classical Approaches to Solve scalability}
The inapplicability of a classical approach to solving the scalability issues of blockchain in cross-cryptocurrency platforms is a major concern. Another challenge lies in understanding the assumptions to be made for achieving trade-off among multiple scalability parameters.

\subsubsection{Batch Processing of Multiple Transactions}
Batch processing refers to reducing the size of transaction record by putting multiple transactions into one. This allows for more transactions per block which causes only one batch payment to be paid for all transactions. Here, instead of complete information of all the transactions, the first transaction information is stored entirely. For the rest of the transactions, only a fraction of the data is stored. This approach helps reduce the amount of transaction data (in bytes) to be stored and the payment cost. But, often, there is a limitation in creating batches of different transactions from different wallets. Moreover, this approach does not provide privacy as details of other transactions in a batch are open.
\subsubsection{Bitcoin Cash – Hard Bitcoin}
Bitcoin Cash is a hard fork of Bitcoin used as an alternative to Bitcoin to add value propositions in terms of transaction speed. The primary motivator of Bitcoin Cash is to increase the default block size of 1 million bytes (as used in Bitcoin) to 8 million bytes, which can speed up the whole transaction processing system and provide a TPS 8x of the default. But, this is a temporary solution to meet the global benchmark and compete in transacting space. Additionally, it might not be adaptable for other blockchain networks.
\subsubsection{Lightning Network for Reserved Bitcoin Transactions}
The lightning network facilitates instant processing of reserved transactions and promotes micro-transacting. The payment channel created on a lightning network does not run transactions through a Bitcoin blockchain rather, it considers them as reserved and charges no additional transaction fee. This approach enables users to take their bitcoins off the blockchain and transact with another party privately. But, the limitations of the lightning network are that it only works for Bitcoin-core-based blockchains like Bitcoin Cash, Litecoin, Digibyte, Dogecoin, etc. \cite{67}
\subsubsection{BloXroute – Blockchain-Agnostic On-Chain Solution}
Unlike off-chain transactions, which take the value outside of the blockchain, an on-chain transaction is considered valid only if the blockchain is modified to reflect the transaction on a public ledger. On the other hand, blockchain-agnosticism refers to a single platform that allows operation from different underlying blockchain technologies. BloXroute is a startup focusing on BDN (Blockchain Distributed Network) while inspired by the concept of CDN (Content Delivery Network). A CDN is a geographically distributed network of proxy servers and their data centers. The main objective of CDN is to deliver content to end-users at a very high speed with the help of a technique called replication. Web content services provided by CDNs are done by duplicating the content from other servers and directing it to users from the nearest data center. Though BDN is in ideation phase, it can give us a future direction to solve blockchain's scalability issue.
\subsubsection{EOS – Scalability through Delegated PoS (Proof-of-Stake)}
EOS is a blockchain project which provides high-end theoretical scalability using delegated PoS. The enormous amount of electricity used in Bitcoin is due to the PoW (Proof-of-Work) algorithm used as a consensus mechanism to secure Bitcoin blockchain. PoW algorithm works by having all the blockchain nodes to solve a cryptographic puzzle. Miners solve this puzzle, and the first one to find the solution gets the miner reward. According to Digiconomist, “Bitcoin miners alone use about 54 Twh of electricity, enough to power 5 million households in the US or even power the entire country of New Zealand.” 

PoW gives more rewards to people with better and more equipment. The higher the hash rate is, the higher the chance to create a block and receive the mining award. Often, mining pools are created by miners where miners combine their hashing power and distribute the reward evenly across everyone in the pool. The above-mentioned issues of PoW open up the requirement of a new consensus mechanism, known as Proof-of-Stake (PoS). PoS uses an election process to randomly choose a node to validate the next block. Choice of validators is not entirely random; instead, a node has to deposit a certain amount of coins as stake into the network to become a validator. There is a linear correlation between the size of the stake and the chances of a validator to forge the next block. Unlike Bitcoin, EOS uses a democratically selected pool of 21 validators to achieve consensus much faster. This does come at the cost of decentralization, as to take control of EOS consensus mechanism, we only need to gain control of more than 50\% of the producers (11 nodes out of 21). Delegated PoS is the key part of delivering the revolutionary speed (millions of TPS) and efficiency that EOS promises to give.

\section{Quantum-aid to Classical Blockchain}
\subsection{Quantum Offers Unprecedented Growth in Computing}
\subsubsection{Qubit} A qubit or quantum bit is the smallest information in quantum computing. It is the quantum version of the classical binary bit, which can be physically realized as a coherent coexistence of both 0 and 1, unlike either of the states in classical computation. Qubit measurement might destroy the coherence and irrevocably cause disturbance in the superposition state. A quantum state of a qubit can be represented by a linear superposition of two orthonormal basis states. A multi-qubit ($n$-qubit) system can be represented by a superposition state vector in $2^n$ dimensional Hilbert space.

\subsubsection{Quantum superposition}
Quantum superposition refers to the simultaneous existence of a quantum particle as mathematical possibilities described by a wave function rather than one actual object. Mathematically, it corresponds to a property of solutions to the Schrodinger's equation where any linear combination of solutions is another solution. In quantum information processing, $\ket{0}$ and $\ket{1}$ are the Dirac notation for the pure quantum states, which will always give the result 0 and 1 respectively when converted to classical logic by a measurement.  Then a quantum state $\alpha \ket{0} + \beta \ket{1}$ will be superposition of linear states $\ket{0}$ and $\ket{1}$ with complex numbers $\alpha$ and $\beta$ as coefficient representing how much $\ket{0}$ and $\ket{1}$ go into each configuration.\cite{ant} This notion of coexistence in multiple states holds true only under the absence of observation.
\subsubsection{Quantum Entanglement}
Even more surprising than superposition, Einstein's theory of "Spooky action at a distance" in EPR (Einstein - Podolsky - Rosen) paradox opens up a new dimension of interpretation of quantum mechanics, known as entanglement. Quantum theory predicts that two entangled
entities might have correlated fates, as the measurement on one entity leads instantaneously to a correlated result. Before the measurement, the outcome of a quantum experiment is unknown, but the correlation between the outcomes prior to obtaining the actual outcome is fully deterministic. Once one outcome of the entangled pair is obtained, there remains no uncertainty about the outcome of the other.

\subsection{Quantum advantage in Blockchain scalability}
Quantum mechanical principles mentioned above can provide substantial speed-up over classical computing. \cite{bwt} Since blockchain platform is rigorously used in multiple fintech applications, we have analyzed four use cases of fintech applications where quantum invasion can be made to provide computational advantage.
\subsubsection{Supercharged data analyses}
Unlocking blockchain's potential to manage IOT devices will help understand the huge volume of data generated and processed by multiple transactions. This analysis primarily helps in decision-making systems where consistency, autonomicity and processing delays can be adjusted. Predictive analytics in Blockchain technology enables trackability of the provenance of items along the supply chain. Lack of sufficient data leads to inaccurate predictions, and available classical systems with classical bits are less efficient in processing a massive volume of data. Exploiting quantum mechanical principles like quantum parallelism can provide a significant speedup in supercharged data analyses. Better utilization of a quantum-enabled analytical model using Blockchain data of a particular organization enables identifying potential return on investment (ROI) sources.
\subsubsection{Greater calculation speed}
Limited block size and difficulty of proof of work of bitcoin consensus protocol lead to a bottleneck in transaction processing capacity. \cite{53} Operations like duplicate record keeping, third-party validations, etc., constitute limited transparency and causes slow data verification. Classical solutions to use a better machine or faster communication links can provide enhanced throughput in transaction processing speed, but to a limited extent (around 5x). Upon the availability of fault-tolerant quantum computers, in the future, the quantum era will provide noiseless qubits for enhanced processing speed, accuracy, and a faster mode of communication using quantum networks. This mode of immediate execution of transactions will incur low computational complexity and reduced resource overhead.
\subsubsection{Reduced False-Positive in fraud detection}
Classical fraud detection models mainly rely on interpreting past payment methods for assigning risk scores and detecting fraud payment transactions. Despite the advent of artificial intelligence and machine learning algorithms, classical intelligent systems cannot adapt themselves in processing real-time dynamic data on per event-by-event basis. This leads to unfiltered false-positive cases, which can drastically affect resource overhead and speed of fraud investigation, resulting in a significant impact on profitability and customer experience. One of the possible solutions to drive down the false positives is the calculation speed and optimality in choosing the data points. This can enhance the accuracy and adaptability of the fraud alert investigations. Quantum machine learning has the potential to achieve significant speedup over classical machine learning models. If we feed more data points into the training model, re-adjustment of fraud detection rules and models to match the perceived level of risk will be enhanced. 
\subsubsection{Efficient Monte-Carlo Simulation}
The cryptocurrency market is still volatile as it is decentralized, open, and highly accessible. Monte Carlo simulation and time series analysis are essential for forecasting future risk and return associated with cryptocurrencies for a selected time period. The classical Monte Carlo algorithm is computationally less efficient under simulation performed over many variables bounded to different constraints. The process incurs huge time overhead, and approximating a solution will involve many computations. If conditions and parameters put into the model are poor and unfiltered, the simulation outcome will be poor. Quantum algorithm can accelerate and achieve quadratic speedup over classical Monte Carlo methods in a very general setting by estimating the expected output value of an arbitrary randomized.

\section{Shor’s Algorithm – Threats on Classical Public-key Blockchain Algorithms}
\subsection{RSA (Rivest Shamir Adleman) Algorithm} In RSA, a user sends its public key to the server and requests some confidential data. The server encrypts the data using users’ public key and sends back the encrypted data. After receiving the data, the user decrypts it using their private key. Blockchains follow a similar algorithm to that of RSA for creating and encrypting the blockchain wallets. \cite{4} Whenever a cryptocurrency wallet is to be created, a public address-private key pair is generated. The public address of the user is used to receive cryptocurrencies and consult respective balances on the blockchain. \cite{16} In contrast, the user's private key is used in correlation to the public key to access and spend the user’s crypto. Mathematically, RSA uses a one-way trapdoor function where public information is two large prime numbers whose product is used as the modulus for encryption and decryption. The intractability lies in the mathematical hardness to calculate back the factors of such a large integer. Until 1994, Peter Shor came up with his novel quantum algorithm for large integer factorization using Quantum Fourier Transform (QFT). A 2048-bit key RSA encryption can be broken by a quantum computer 4096 logical qubits.\cite{4}
\subsection{ECC and ECDSA (Elliptic Curve Cryptography and Elliptic Curve Digital Signature Algorithm)} Elliptic curve-based algorithms are used as standards for creating keys for wallet and signing for transactions. With ECDSA, a random 256-bit private key is generated to derive the public key.\cite{5} \cite{38} Classically, it is very difficult to find the random value for a given generated public key. The private key generated randomly is used to sign for transactions, and the public key is used to prove that a private key has been used. But, Shor’s algorithm can reverse back the private key for any public address and gain control over a person’s wallet. Though from a cryptographic primitive “security-level” (in terms of bits or number of operations an attacker has to perform to compromise the security of the underlying system) point of view, ECDSA provides a better security level than RSA. A 224-bit sized public key ECDSA provides a 112-bit security level. On the contrary, to achieve the same 112-bit security level, a 2048-bit sized public key is required in RSA. But, with respect to post-quantum resistance, ECDSA is under more threat than conventional RSA.\cite{4}\cite{6}

According to public research, “RSA 2048-bit keys require around 4096 qubits (5.2 trillion tofolli quantum gates) to be defeated, whereas ECDSA 256-bit keys require only 2330 qubits (126 billion tofolli quantum gates).” \cite{5}

\subsection{ECDH (Elliptic Curve Diffe Hellman) Algorithm}
ECDH is a key agreement protocol to allow two parties to establish a shared secret over an insecure channel. This concept is used in the context of blockchain by having the bitcoin receiver publish some ECDH-information which is to be used by the sender for calculating a shared secret.\cite{7} This shared secret is the bitcoin address, also referred to as a stealth address, reusable payment codes, reusable address, or paynyms to which a sender sends his money. \cite{7} The receiver calculates the corresponding private key to gain access to the money. The underlying mathematics for key exchange protocol relies on the Discrete Logarithm Problem (DLT). The objective is to determine a unique integer that can be the order of the generator element of a finite cyclic group. For general elliptic curves, DLP seems to be extremely computationally hard as best known classical algorithms for DLP on elliptic curves are the generic algorithms with running times exponential to the number of bits necessary to describe the problem. Modified Shor’s algorithm for DLP can theoretically achieve exponential speedup to break the ‘thought to be’ irreversible and intractable one-way function.
\subsection{DSA (Digital Signature Algorithm)}
In order to guarantee non-repudiation (post-denial) of information in the blockchain system, digital signatures are used to verify the integrity of a file or a message. But, lack of diversity in industrial settings makes currently available cryptosystems susceptible to quantum computer-aided attacks. The degree of security provided by Classical Digital Signature (CDS) schemes rely on intractability and complexity of large integer factorization and discrete logarithmic problem. This makes the system to be no longer quantum-safe in the near-term era.\cite{8}

\section{Grover’s Algorithm – Threats on Classical Blockchain Techniques}
\subsection{Faster Detection of Hash Collision} Blockchain relies on the computation of hash functions to provide security against modification of past blocks and thus guarantees the integrity of blocks. Underlying distributed architecture demands massive computational effort in recomputing a chain of blocks, finding a hash collision with existing hash and modifying a single block are of extreme challenge.\cite{19} But Grover’s algorithm can specifically provide a solution to the problem of finding the pre-image of a value of a function that was initially difficult to invert. \cite{13}\cite{14}Given a hash value (signature) and a hash function as input to the quantum computer, Grover’s algorithm can give us the original data on which the hash was applied in quadratically smaller number of efforts. This allows more efficient and accelerated generation of hash collisions rather than brute-force search. \cite{10} Hence, it offers achievable speedup in detection of hash collisions which can be used to replace blocks without disturbing the integrity of blockchain – which can be a serious security threat to classical blockchain.
\subsubsection{Recreation of entire blockchain}
Grover’s can speed up the generation of nonces to an extent where the entire chain of records can be recreated with consistently modified hashes. A classical attack necessitates using linear-time brute-force search, which involves the whole hash space until a match is found with the known hash value. \cite{14} This potentially causes vulnerabilities in the system as inserting a modified block will remain unspotted as it will not compromise the sequential consistency of the blocks in the blockchain. 

\section{Post- Quantum Initiatives for quantum-resistant cryptography}
Classical cryptographic primitives used in blockchain technologies are under threat due to quantum computers. Pre-quantum and post-quantum security level of different classical cryptographic algorithms are shown in tables \ref{table:2} and \ref{table:3} In order to overcome the aforementioned cryptographic challenges in classical blockchain, several post-quantum cryptographic initiatives have already been taken. \cite{maindoc2} The main aim behind these initiatives and their relevance with respect to classical security systems are specified as follows.

\begin{center}
\begin{table}
\begin{tabular}{p{1.5cm}|p{1cm}|p{1cm}|p{1.2cm}|p{6.5cm}}
\hline
Classical \newline Algorithm & 
Current \newline security \newline level (bits) & 
Post-quantum \newline security \newline level (bits) & 
Quantum \newline algorithm \newline causing threat & DLTs under threat\\
\hline
\hline
SHA-256 \newline (Secure Hash \newline Algorithm) & 256 & 128 & Grover & Bitcoin, Ethereum, Dash, Monero, Litecoin, Zeash, Ripple, NXT, Byteball \cite{52}\\
\hline
Keccak-256 & 256 & 128 & Grover & Ethereum, Monero, Bytecoin \\
\hline
Keccak-384 & 384 & 192 & Grover & IOTA \cite{18} \\
\hline
Keccak-512 & 512 & 256 & Grover & Ethereum \\
\hline
Scrypt & 256 & 128 & Grover & Litecoin, NXT \\
\hline
RIPEMD 160 \newline (RIPE Message Digest) & 160 & 80 & Grover & Bitcoin, Ethereum, Monero, Litecoin, Ripple, Bytecoin \\
\hline
SHA-3 256 & 256 & 128 & Grover & - \\
\hline

\end{tabular}
\caption{Hash-function based algorithms under quantum threat}
\label{table:2}
\end{table}
\end{center}

\begin{center}
\begin{table}
\begin{tabular}{p{1.5cm}|p{1cm}|p{1cm}|p{1.2cm}|p{6.5cm}}
\hline
Classical \newline Algorithm & 
Current \newline security \newline level (bits)& 
Post-quantum \newline security \newline level (bits) & 
Quantum \newline algorithm \newline causing threat & DLTs under threat\\
\hline
\hline
ECDSA & 128 & Broken & Shor & Bitcoin, Ethereum, Dash, Litecoin, Zeash, Ripple, Byteball \\
\hline
RSA-1024 & 80 & Broken & Shor & - \\
\hline
RSA-2048 & 112 & Broken & Shor & - \\
\hline
RSA-3072 & 128 & Broken & Shor & - \\
\hline
AES-128 & 128 & 64 & Grover & - \\
\hline
AES-256 & 256 & 128 & Grover & - \\
\hline
\end{tabular}
\caption{Asymmetric and symmetric encryption algorithms under quantum threat}
\label{table:3}
\end{table}
\end{center}

\subsection{PQCrypto} This is one of the initiatives to allow users to switch from classical cryptography to post-quantum cryptography so that they remain secure for the long term against attacks from quantum computers. The primary aim of PQ Crypto is to design a portfolio of post-quantum public-key cryptosystems with (i) improved speed, (ii) enhanced security, (iii) adaptability to several performance challenges of IoT, Cloud, and mobile devices, etc.\cite{20}
\subsection{SAFEcrypto}
The word SAFEcrypto stands for Secure Architectures of Future Emerging Cryptography. Its primary aim is to provide robust, physically secure solutions for future IOT applications in the quantum era. \cite{21} Underlying mathematical source to provide computational hardness is based on lattice-based cryptographic schemes to improve the security of digital signatures, authentication schemes, public key encryption, attribute-based encryption (ABE) and identity-based encryption (IBE), etc. Since post-quantum initiatives require low power consumption and real-time performance, SAFEcrypto aims to achieve 10x speed up throughout for real-time application scenarios and 5x reduction in energy consumption for low power and embedded mobile applications. The algorithmic optimization and improved design using lattice-based cryptographic schemes can enhance the performance of both resource-constrained and real-time internet applications.
\subsection{CryptoMathCREST}
This post-quantum initiative is a part of the CREST funding program, which is run by the Japan Science and Technology Agency (JST). The main aim behind this initiative is to study the mathematical problems and their variations underlying the security modeling of a next-generation cryptosystem.
\begin{figure}
  \includegraphics[width=12cm , height=7cm]{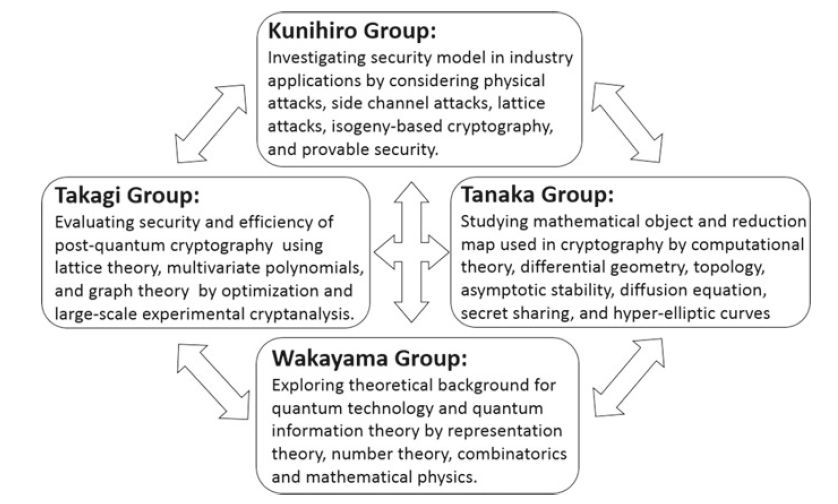}
  \caption{Key research areas of CryptomathCREST post quantum initiative \cite{pic3}}
  \label{fig:research-group}
\end{figure}
The novel mathematical techniques emerged as research outcome will serve as the basis of the construction of new cryptographic protocols. The four key research areas of CREST Crypto Math project comprising of 25 mathematicians and four postdocs involve investigation of the security model in industry applications, evaluation of security and efficiency of post-quantum cryptography, study on mathematical object and reduction map used in cryptography, and exploration of theoretical background for quantum technology and quantum information theory as shown in figure \ref{fig:research-group}.

\subsection{Post Quantum Standardization}
\begin{figure}
  \includegraphics[width=12cm , height=6cm]{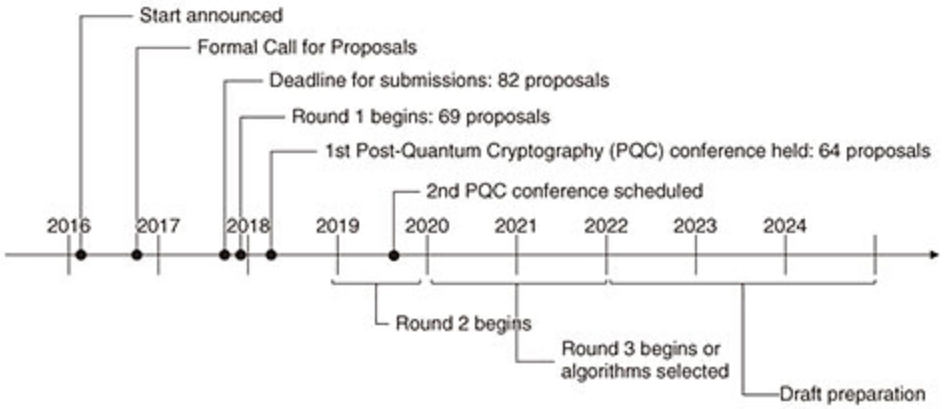}
  \caption{NIST standardization project timeline}
  \label{fig:nist-time}
\end{figure}

Post quantum cryptography standardization is a program hosted by NIST to update their standards where Post-quantum cryptography (PQC) or quantum resistant algorithms (QRAs) can be incorporated. Symmetric key cryptographic primitives like Advanced Encryption Standard (AES), Data Encryption Standard (DES) are comparatively easy to modify in a way that helps them increase their security level (in terms of bits) and become quantum resistant. \cite{22} So, the primary focus of the NIST PQC initiative is on public-key cryptographic primitives like digital signature and Key encapsulation mechanisms (KEMs), currently, NIST initiated a call for proposals is in round three out of the expected four rounds. The call for proposals and timeline of NIST Post Quantum Cryptography Standardization project has been depicted in figure \ref{fig:nist-time}. \cite{23}\cite{24}\cite{33} These post-quantum cryptographic initiatives can have a diverse range of security applications. \cite{25}\cite{27}\cite{29} The more precise post-quantum initiative designed for blockchain includes Bitcoin Post Quantum, Ethereum 3.0, and quantum-resistant cryptosystem for Abelian platform and PQC for Corda Blockchain to name a significant one.\cite{28} In the next section, we will discuss all these four post-quantum initiatives, including their operational principle, pros and cons, relevance with respect to current Blockchain technology, and underlying challenges in implementation. \cite{26}\cite{30}

The timeline for NIST standardization project is summarized below. \cite{31}\cite{32}
\begin{itemize}
\item February 2016: Announcement of the start of the Post-Quantum Cryptography Standardization Project
\item August 2016: Release of NISTIR 8105, Report on Post-Quantum Cryptography
\item August 2016: RFCs on Submission Requirements and Evaluation Criteria
\item December 2016: Formal Call for Proposals
\item November 2017: Deadline for submissions
\item December 2017: Examination of documents and forms; Round 1 begins
\item April 2018: First Post-Quantum Cryptography (PQC) Standardization Conference
\item 2018/2019: Round 2 begins
\item August 2019: Second PQC Standardization Conference (plan)
\item 2020/2021: Round 3 begins or algorithms to be selected
\item 2022/2024: Draft preparation to be completed
\end{itemize}

\section{Post Quantum Initiatives on Blockchain}
\subsection{Bitcoin Post-Quantum}
\subsubsection{Bitcoin Security Primitive/ Operational Principle} 
The security of decentralized digital cryptocurrency, Bitcoin, relies upon the ECDSA and Proof of work (PoW) algorithm to authorize the payer and prevent a payer from double-spending. \cite{3} Transfer of bitcoins between users happens through transactions from one address to another, where address represents the hash of the receiver’s public key, which is revealed when distributing the transaction over the bitcoin network.\cite{6} The amount of bitcoin ownership depends on the private key ownership corresponding to the given public key. The digital signature used to sign the transaction shows the proof of ownership.
\subsubsection{Quantum- threat to Bitcoin security}
ECDSA strength is based on a discrete logarithm problem that is vulnerable to Shor’s algorithm. \cite{9} The corresponding private key can be calculated using a quantum computer whenever a public key is disclosed in the network. A new transaction can be forged with a valid digital signature. But, the process of searching the pre-image of the hash is still computationally complex, even in a quantum computer. Quadratic speedup of Grover’s algorithm will not put significant speedup on hash functions of appropriate length.
\subsubsection{Quantum – safe hash-based cryptography}
For PQC era, the extended Merkle Signature Scheme (XMSS) can be used as a quantum-safe digital signature scheme because of its high security, acceptable signature length, and permissible key generation time. \cite{39} To achieve a post-quantum security level, at least 256-bit hashes (as used in P2WSH, Pay-to-Witness-Script-Hash) must be to provide 128-bit collision resistance. XMSS digital signature scheme combined with RFC 8391 can naturally resist side-channel attacks and guarantees security against quantum attacks by providing 128-bit post-quantum security level (256-bit hash). \cite{17}
(iv) Usability and relevance to current blockchain:
In order to find an authentication path from a given XMSS tree, the required number of signatures is an exponential factor of the height of the tree. An increase in the height of the XMSS tree will imply more transactions to be signed (2h= 1024, 1048576 for h=10, 20 respectively). On a modern classical computer, this can consume time ranging from less than a second (h=10) to about 10 minutes (h=20) despite being quantum-safe. \cite{42}\cite{43} Moreover, as per Bitcoin Post – Quantum (BPQ) consensus, the most prominent approach is to generate both XMSS key and a quantum-safe address. However, BPQ approach discovered so far is impractical for Blockchain applications as XMSS is computationally slow. \cite{11} \cite{34}

\subsection{Ethereum 3.0}
\subsubsection{Ethereum Security Primitive/ Operational Principle}
Ethereum uses zero-knowledge (ZK) proof technologies, enabling one party to prove his knowledge to another party without actually conveying the information itself.\cite{58} - \cite{61} This privacy-enhancing technique allows Ethereum to be a scaling technology as the amount of information transfer between users will be significantly reduced, and proofs can be verified at a much faster rate. \cite{58} \cite{59}
\subsubsection{Quantum threat to Ethereum Security}
Ethereum cryptocurrency relies on ZK-SNARK (Zero-Knowledge Succinct Non – Interactive Argument of Knowledge) technique. \cite{50} \cite{51} Though this technique is non – interactive, meaning code can be run or deployed autonomously, it requires a trusted setup. Trusted setup architecture requires key creation to generate and verify proofs for private transactions. The secrets which are used to generate these keys can be used to forge transactions unless destroyed. The underlying cryptographic primitive used for ZK-SNARK is another disadvantage of Ethereum technology as it primarily relies on elliptic curve cryptography. This can put threats on Ethereum security as finding the discrete logarithm of the random elliptic curve for a given publicly known base point will not remain infeasible upon the availability of a full-scale quantum computer.
\subsubsection{Quantum – safe cryptographic primitive (ZK - STARKS)}
ZK-STARK (Zero-Knowledge Scalable Transparent Argument of Knowledge) is the post-quantum initiative of Ethereum 3.0, where post-quantum security can be achieved using quantum-resistant hash functions.\cite{13}\cite{35} Additionally, the requirement of trusted set–up architecture is no longer there in the utilization of ZK-STARK in the network. This helps to remove the vulnerability of the trusted parties compromising overall system privacy. The advantage of STARK can be realized on improved scalability and enhanced privacy. \cite{51} It can perform bulk computations and storage off–chain allowing exponential scaleup of blockchain infrastructure without losing computational integrity.
\subsubsection{Usability and relevance to current blockchain}
Although issues like the requirement of trusted setup architecture and quantum attacks on elliptic curve algorithms, technology like ZK – SNARK has a faster adoption rate over quantum-resistant ZK – STARKS. \cite{17} Reasons behind the lower acceptance rate of ZK – STARKS are lack of developer libraries, published code, projects, and developers currently available to work on that technology.  Moreover, the development of quantum computers in terms of more availability of noise-free qubits to perform those quantum attacks is also in its nascency and far costlier for the end-user concerning the current scenario.
\subsection{Abelian}
\subsubsection{Blockchain Privacy Primitive/ Operational Principle}
Data privacy has always been one of the major challenges in near–term blockchain applications. Cryptographic tools can preserve privacy in the case of both permission-based and permission-less applications. The applications like Fully Homomorphic Encryption (FHE), Multiparty Computation (MPC), Functional Encryption (FE), Searchable Encryption (SE) have several challenges and computational complexities associated with them. \cite{36} FHE seems to be the universal solution provider to any blockchain privacy issue but is not implementable in the near future. MPC technique is highly complex and inefficient in terms of computation and communication though they can provide privacy and verifiability in distributed computing platforms. FE and SE techniques lack theoretical research to devise compelling use cases and more efficient schemes for large datasets without compromising security. 
\subsubsection{Privacy Perspective of Major Cryptocurrencies}
The existing cryptocurrencies can provide dual privacy due to their ability to keep anonymity of originator and recipient of each transaction while simultaneously hiding each transaction amount. But these techniques cannot offer any solution that can be quantum-safe and, at the same time, can provide privacy with accountability. The privacy perspective of major cryptocurrencies has been shown in table \ref{table:1}.

\begin{center}
\begin{table}
\begin{tabular}{l|p{10cm}}
  \hline
  Bitcoin & (a) Pseudonym from privacy concern.
 \newline (b) Wallet addresses and personal information of wallet holders are kept independently.
 \newline (c) Basic privacy level.
 \newline (d) Transactions are traceable as all transactions are recorded on the public ledger.
 \newline (e) Threat: if a graph can be built to represent all wallet addresses as nodes and all transactions as edges, the complete view of the bitcoin system can be obtained from a single bitcoin full node through graph analytics.  \\
  \hline
  Dash & (a) Improvement over Bitcoin using privacy-centric master nodes to weaker the traceability.
 \newline (b) Uses ‘mixing technique’ to obscure the linkage between inputs and outputs of individual transactions by combining them with inputs and outputs of other transactions.
 \\
  \hline
    CryptoNote & (a) Improvement over both Bitcoin and Dash.
\newline (b) Uses ring signature to hide the identity of the originator.
\newline (c) Applies Diffe – Hellman key exchange protocol to hide recipient’s identity.
 \\
  \hline
   Zerocoin & (a) Breaks Linkage between individual transactions to improve privacy. \\
  \hline
    Monero & (a) Combines techniques from both Dash and CryptoNote to hide each transaction amount.
\newline (b) Applies Zero – knowledge range proof to ensure that the originator cannot overspend any amount.
\newline (c) One of the cryptocurrency techniques considered to be providing full privacy.
 \\
  \hline
   Zerocash & (a) Breaks linkage between transactions
\newline (b) Can hide the identity of the user, thus providing anonymity.
\newline (c) Applies Zero-Knowledge SNARKs to cover up each transaction amount.
\newline (d) Another cryptocurrency technique to provide full privacy.
 \\
  \hline

\end{tabular}
\caption{Privacy Perspective of Major Cryptocurrencies}
\label{table:1}
\end{table}
\end{center}

\subsubsection{Quantum – safe Abelian Cryptocoin (ABE)}
ABE is designed to build an efficient and quantum-safe privacy coin, which uses lattice-based cryptographic constructions to provide accountability. Users of ABE are allowed to choose the privacy level for each of their transactions. Its basic privacy level is comparable to bitcoin, while the full privacy level is compared to Monero or Zerocash. Along with these two privacy levels, ABE allows users to pick a designated authority who can link the current transaction with the immediately previous transaction and simultaneously crack the actual transaction amount. Cryptographic primitives of ABE are based on the intractability of lattice-based hard problems, namely Short Integer Solutions (SIS) and Learning with Errors (LWE). These schemes used in ABE are supported by rigorous proofs of post-quantum security, thus providing a practical and scalable solution with an enhanced privacy perspective.
\subsection{Corda}
Corda is probably the most cryptographically agile DLT technique that allows users to choose between multiple signing key types. \cite{37} These key-types include both the techniques which are (a) not quantum-safe like RSA, ECDSA, EdDSA (Edwards-curve Digital Signature Algorithm), and (b) quantum-safe like SPHINCS (post-quantum stateless hash-based signature scheme).\cite{77}\cite{4} SPHINCS-256 is a high-security scheme that can sign hundreds of messages per second on a modern 4-core 3.5 GHz Intel CPU.\cite{5}\cite{6} SPHINCS provides signatures of size 41 KB and both public keys, private keys of size 1 KB each. Conventional hash-based signature schemes are stateful, meaning they need to record information (state) after processing every signature. \cite{40} Generally, input to these hash-based signature schemes is a secret key, and the message itself generates a signature and an updated secret key as output. \cite{14} In case there is any failure while updating or copying the key from one device to another, the system's overall security is susceptible to disintegration. Unlike these schemes, SPHINCS is a method to apply randomized tree-based stateless signatures. This can provide long-term $2^{128}$ security, which is resistant to any quantum attack in the near term future. Moreover, this signature scheme is also practical as it can provide a good trade-off between speed and signature size.

\section{Feasibility of Quantum Attack on Blockchain DLTs}
The underlying dependencies of the cryptographic schemes and lack of quantum-resistant algorithms make current Blockchain DLTs vulnerable to several attacks made by quantum computers. Hence, it is necessary to understand the current level of vulnerability that can be exploited upon availability of near-term quantum computers and the required number of computational metrics (in terms of qubit count, gate fidelity, qubit noise, underlying physical machine description) to execute those quantum attacks.\cite {2dq}\cite{qdlc} The following section will describe the mathematical and operational feasibility of several quantum attacks on Blockchain technologies like Bitcoin, Ethereum, Litecoin, Monero, and Zcash.

\subsection{Bitcoin}
Bitcoin incorporates Hashcash as its PoW, which was designed as a countermeasure to Denial of Service (DoS) for email systems. This enables the potential sender to send an email only after solving a computationally hard problem. In Hashcash, the prospective miner must calculate an SHA-256 hash value for the header and a random number. This random number is a significant parameter to measure the computational difficulty of the hard problem to be solved. The transaction mechanism of Bitcoin incorporates an ECDSA signature scheme to prove authority, ownership, transaction immutability, and guaranteed non-repudiation on token usage. The Elliptic curve being employed is secp256-k1. \cite{15} \cite{64} \cite{65} This signature is made up of two values, S and R, where R denotes x coordinate of a point on the elliptic curve and the corresponding another half of the signature (S) can be obtained as follows:
\newline
$S=K-1 ( SHA256 (M) + dA.R )mod\ p$ where,
\newline
K = temporary private key.
\newline
dA = signing private key.
\newline
p = prime order of the elliptic field.
\newline
SHA256 (M) = Output of the message M when SHA-256 hash algorithm is applied. \cite{63}
\newline
With its quadratic speed up, Grover's algorithm can perform PoW at a much faster rate than classical miners. If an attacker can generate as much as PoW as the rest of the network combined (>50\%), it can allow an attacker to enforce a consensus on any block of the network.
Another source of vulnerability is the use of ECDSA scheme for transactions. A classical computer can solve the hardness of the ECDSA scheme in $O(2n)$ complexity, which provides computational overhead. Shor’s algorithm can perform the same computation exponentially faster than the classical counterpart. \cite{9} The hard mathematical problem underlying ECDSA is the discrete logarithm problem, which Shor’s algorithm can solve on a quantum computer in $n^3 log (n) log log (n)$ time, which is of $O (n^3)$ complexity. Moreover, an effective quantum attack can be made on Bitcoin when a public key and a signed transaction have already been broadcasted to the network, and the objective is to find the corresponding private key. A quantum computer can speed up the process of signing a new transaction with the private key. This will help an illegitimate user to impersonate as the key owner and place that illegitimate transaction before the original transaction on the blockchain. Thus, the original transaction can be stolen, and the newly created Unspent Transaction Output (UTXO) can be directed into an account, thus leading to the existence of a parallel blockchain. It has been experimentally shown that a quantum computer with 4,85,550 qubits running at a clock speed of 10GHz can solve the problem in 30min. \cite{62}

\subsection{Ethereum}
 Ethereum is primarily designed to introduce the use of smart contracts and distributed applications where each transaction will add or deduct Ether to a user’s account. The earlier consensus mechanism of Ethereum is Ethtlash, a PoW where only one round of SHA-3 hashing is used. \cite{11} The power of a miner is determined by the amount of stake held by a miner in terms of Ether. Like Bitcoin, Ethereum uses a variant of ECDSA where the primary public key associated with a user account is not revealed to the network. Instead, the public key can be retrieved from another user’s transaction signature using the public key recovery technique. EthHash is vulnerable to be attacked by a quantum computer due to Grover’s algorithm. The difficulty value for a successful attack on Ethereum can be described as follows:
\newline
$D = (H_r * B)/2^{32}$, where
\newline
D = Difficulty of the PoW mechanism 
\newline
$H_r$ = Network hash rate
\newline
B = Block time of the Blockchain
\newline
In classical system, for B=16sec and $H_r=18*10^{3}$  H/sec, the difficulty value D=670552. On the other hand, the equivalent hash rate on a quantum computer with operating frequency is $0.04 * S\sqrt{D}$. \cite{66} This enables a quantum computer to perform a 51\% attack with only a clock speed of 5THz. Ethereum is also vulnerable to Shor’s algorithm due to its similarity with ECDSA. Unlike Bitcoin, Ethereum has a significantly shorter transaction processing time (TPS) which is advantageous. But, Ethereum is significantly more vulnerable to quantum attack due to its account-based transaction system. In Ethereum, all outgoing transactions have to be signed using a single public key/private key pair associated with the account. A quantum attacker can request the public key and calculate the private key using Shor’s algorithm exponentially faster. \cite{9} This can make the entire account balance vulnerable after a single transaction.

\subsection{Litecoin}
 Litecoin is designed to process transactions much faster than Bitcoin. To increase TPS, Litecoin uses a different PoW scheme called Scrypt, which can consume significantly less hashing power than Bitcoin. \cite{55}\cite{56} (Hashing rate of approximately 46*106 TH/sec for Bitcoin and 298TH/sec for Litecoin). \cite{68} \cite{69} Thus, Litecoin focuses on a highly intensive use of RAM on mining nodes rather than the highly intensive processing power. \cite{54} But, since Litecoin is a source-code fork of Bitcoin Blockchain, it is also potentially vulnerable to quantum attacks. The signature scheme used in Litecoin is also secp-256k1 elliptic curve of ECDSA, which makes it vulnerable to Shor’s algorithm. \cite{15} Moreover, the PoW of Scrypt is also vulnerable to be attacked by a quantum computer due to Grover’s algorithm. The difficulty value for a successful attack on Litecoin can be described as follows:
\newline
$D = ( 32*10^{13}*150)/2^{32}= 11175870$

Litecoin’s current hash rate being 320 TH/sec allows a quantum computer to attack the current hash rate while operating with a clock speed of 2.4 THz. \cite{69} Litecoin has a slight advantage in resisting quantum attacks due to shorter block time and slightly faster throughput over Bitcoin. But a decreased hash rate due to reduction of the block reward to complete given PoW may cause the system more vulnerable and susceptible to quantum attack.

\subsection{Monero} 
Monero is a Blockchain to provide user’s privacy by obfuscating both user’s identity and the amount of transaction through Pedersen Commitments and Range Proofs. CryptoNight V8 PoW scheme of Monero is memory – intensive process that makes access to slow memory at random intervals. \cite{71} Monero uses EdDSA as a signing algorithm implemented using Twisted Edwards curve Ed25519. \cite{75} Like ECDSA, this signature scheme also relies on the mathematical hardness of discrete logarithm problems for its security and thus can be thought of to be vulnerable to quantum attacks by Shor’s algorithm. \cite{9} Moreover, Monero is able to provide transaction anonymity through the use of stealth addresses, ring signatures, and ring confidential transactions. \cite{45}-\cite{49} Bulletproof is the current zero-knowledge proof mechanism used in Monero to ensure the balance of transactions. The use of this mechanism helps Monero be efficient in terms of less computational overhead and less space required on the blockchain. \cite{72} Recently, Monero has moved its PoW scheme from CryptoNight V8 to RandomX. \cite{73} RandomX is designed to execute random programs in a special instruction set comprising of integer arithmetic, floating-point arithmetic, and branch instructions. Monero is slightly more resilient to quantum attacks as the attacker can not obtain information about the amount being transferred in a target transaction. This privacy provides additional security to Monero by making it less attractive security to Monero by making it less attractive for quantum attacks. It should be noted further that RandomX is still thought to be safe from quantum attacks as no known vulnerabilities have been obtained so far.

\subsection{Zcash}
Zcash is a privacy-based Blockchain that allows transactions between both public and private accounts. Zcash transactions rely on ZK – SNARK works on trusted set – up architecture to demonstrate the fairness of a transaction being sent. Equihash is the PoW scheme used in Zcash, which is based on a generalized birthday problem. \cite{70} Like Monero, Zcash is also implemented using EdDSA as a signing algorithm and employing elliptic curve Ed25519. \cite{15} \cite{57} \cite{75} This signature scheme also relies on the hardness of discrete logarithm problems and thus making it vulnerable to quantum attack by Shor’s algorithm.
Moreover, there is a quantum algorithm developed by Grassi et al. that can solve the generalized birthday problem or the K-xor problem in $O(2^{n/(2 + log2k )})$ compared to the best known classical algorithm known as Wagner’s algorithm, which can solve the same problem in $O(2^{n/(1 + log2k )})$. \cite{74} This shows a quantum advantage in terms of reduced time and space complexity and, consequently, can lead to a quantum 51\% attack against the system. Quantum attacks can be made to create an infinite number of Zcash tokens if the attacker can possess the global private key using Shor’s algorithm. If tokens can be created at will and the token creation can be kept unknown to the rest of the network through obfuscation, the overall security and the network performance can be drastically affected. Hence, it can be concluded that Zcash is more vulnerable than existing blockchain technologies due to its low resilience to quantum computing attacks.

\section{Realization of Post – Quantum Block and Future Research Direction}
Post-quantum blockchain techniques do not use quantum mechanical principles to exploit computational speedup. Rather, they are advancements made on classical computation so that current cryptographic primitives used in blockchain can be made resistant to future quantum computing attacks. Researchers have been putting a lot of effort into discovering new post-quantum schemes that can be realized in the near-term future and provide a practical, scalable solution for implementation. This transition of blockchain from pre-quantum to post-quantum era needs to attain certain objectives for their successful integration. Primary challenges involved while devising any new research in this field can be described as follows.

\subsection{Key size and signature size}
Post-quantum cryptosystems claim to offer more security level over existing classical standpoint. Enhancement of security level can be achieved by increasing the number of bits used in cryptographic keys and signatures, which can drastically affect the overall performance of the Blockchain network. To achieve a 128 – bit quantum security level, required key sizes are 2,688 – bit for a public key, 384 – bit for a private key, and 120 KB for the signature. This would necessarily require the storage of a massive amount of information and thus incorporating huge computational overhead. Hence, research is required to devise new techniques to provide a trade-off between achieved security level and Blockchain system performance.

\subsection{Key Generation rate}
Pre–quantum Blockchain schemes allow multiple messages to be signed with the same key. But, if the security level has to be increased, it is necessary to limit the number of messages which can be signed with the same key. Generating new keys with a high speed relies on pseudo-random number generators (PRNGs), which depend on the mathematical properties of a complex algorithm to generate the keys. These keys do not provide enough entropy and are not inherently random, leading to statistically correlated keys. An alternative approach relies on true hardware-based random number generators (HRNGs), which can generate true random bits using properties of classical physics. Though these hardware-based RNGs offer sufficient entropy, their key generation rate is not sufficient for an increase in demand. Recently, researchers are working on quantum-based RNGs, which can exploit the properties of quantum mechanical phenomena to generate true random bits. Thorough research is required to focus on the trade-off between key generation mechanism and rate of key generation for improved efficiency of the blockchain. 

\subsection{Blockchain compliance issues}
Multiple post-quantum Blockchain initiatives have been taken, and the research is still in progress. Hence, constant monitoring of the standard post-quantum schemes that are being approved should be made by the researchers and developers to avoid any compliance issues. 

\subsection{Incompatibility issue with existing hardware}
Post-quantum cryptosystems are highly computationally intensive, as they incorporate large computation overhead to increase system security. But, this overhead requires a huge computational resource with increased processing power. This makes the post-quantum schemes less scalable and hence, less practical to be implemented. Research is required to resolve this classical hardware incompatibility so that a trade-off can be obtained between overall security and computational complexity.

\subsection{Ciphertext overhead}
\begin{figure}
  \includegraphics[width=12cm , height=11cm]{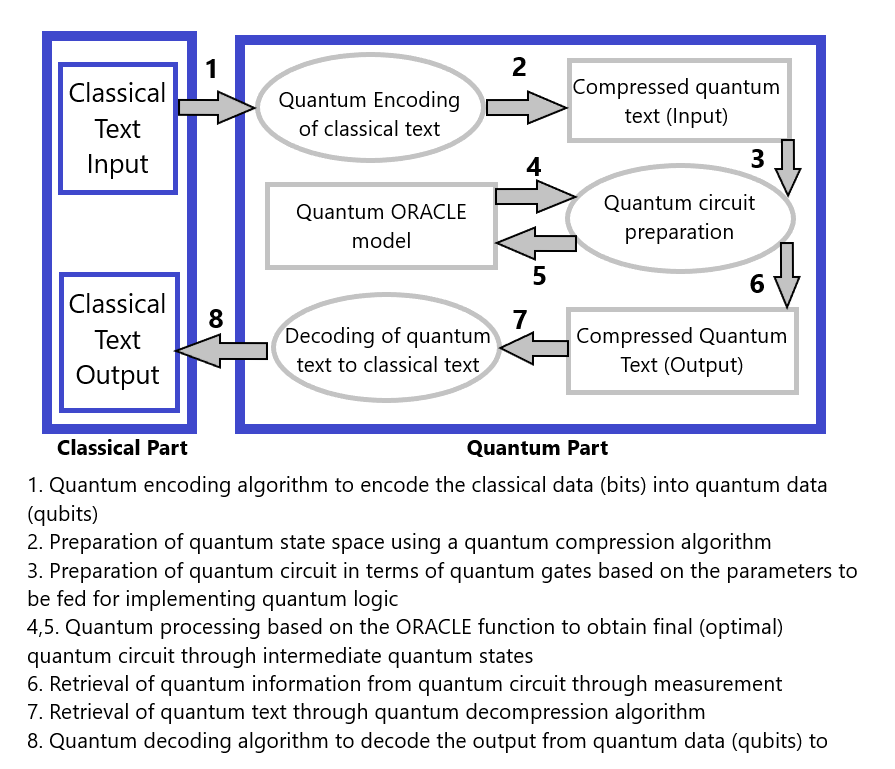}
  \caption{Proposed Block Diagram for Quantum-enabled Compression Technique}
  \label{fig:qcipher}
\end{figure}

Post-quantum cryptosystems are supposed to generate large Ciphertext so that their deciphering can be made complex. But, this puts a lot of overhead on the overall performance of the resource-constrained embedded systems. Quantum computing can be one of the potential solutions to reduce the Ciphertext overhead. Quantum parallelism through quantum superposition can be exploited to prepare a quantum state space where classical data can be encoded in qubits rather than in bits as shown in figure \ref{fig:qcipher}. However, this brings another challenge: segregating the quantum computation part from the classical computational part. \cite{qdlc} If a quantum hybrid system can be incorporated in the post-quantum blockchain, it can aid overall system improvements in terms of security and computational speed.
\section{Conclusion}
Despite the nascency of technological advancements in quantum computing, it has attracted researchers and developers to keep their eye on potential advantages and threats by quantum computers in blockchain and other DLTs. In this article, we have discussed the existing and future scalability aspects of classical Blockchain technologies. The potential benefits to serve blockchain–aided fintech applications offered by quantum computing have been analyzed so that use cases of Blockchain performance in real-world applications can be enhanced further. We have analyzed the impact of Grover’s algorithm and Shor’s algorithm in posing threats to classical Blockchain security primitives. \cite{10} For this purpose, the most relevant Blockchain DLTs like Bitcoin, Ethereum, Litecoin, Monero, Zcash have been reviewed. Their areas of vulnerabilities to be attacked by real quantum computers have been analyzed in detail. In addition, post-quantum initiatives for quantum-resistant cryptography and specific post-quantum initiatives on blockchain have been studied with respect to their operational principle, possible quantum threat, quantum-safe cryptographic primitive, and relevance of those schemes in the current Blockchain scenario. The challenges and future research direction in realizing post-quantum blockchain and exploiting quantum technology have been added in this article. Thus, the article can provide useful guidance to the research areas of quantum computing and blockchain technology and, moreover, in understanding a few possible quantum invasions in post-quantum blockchain to devise new quantum-resistant Blockchain techniques.

\end{document}